\documentclass[final]{IEEEtran}
\usepackage[dvips]{graphics}
\usepackage{epsf,rotating,color,graphicx,float,amssymb,amsmath}
\usepackage{epsfig,url,times,cite,subfigure,soul}
\usepackage{booktabs}
\usepackage{multirow,multicol}
\usepackage{lipsum}
\usepackage{algorithm,algorithmic}
\usepackage{nomencl}
\usepackage{epstopdf}

\usepackage{calc}
\newenvironment{notationlist}[1]
{\begin{list}{}%
{\renewcommand\makelabel[1]{##1\hfill}%
\settowidth\labelwidth{\makelabel{#1}}%
\setlength\leftmargin{\labelwidth+\labelsep}
\setlength\itemsep{0\baselineskip}}}%
{\end{list}}

\begin{document}

\title{\Large  Load Forecasting Based Distribution System Network Reconfiguration \\ --A Distributed Data-Driven Approach}

\author{\large Yi Gu$^1$, \textit{Student Member}, IEEE, Huaiguang Jiang$^2$, \textit{Member}, IEEE, Jun Jason Zhang$^1$, \textit{Senior Member}, IEEE, Yingchen Zhang$^2$, \textit{Senior Member}, IEEE, Eduard Muljadi$^2$, \textit{Fellow}, IEEE, and Francisco J. Solis$^3$, 

{\normalsize $^1$Dept.~of Electrical and Computer Engineering, University of Denver, Denver, CO, 80210\\
$^2$National Renewable Energy Laboratory, Golden, CO 80401\\
$^3$School of Mathematical and Natural Sciences, Arizona State University, Glendale, AZ, 85306}
\vspace{-0.3in}

}

\date{}
\maketitle

\begin{abstract} In this paper, a short-term load forecasting approach based network reconfiguration is proposed in a parallel manner. Specifically, a support vector regression (SVR) based short-term load forecasting approach is designed to provide an accurate load prediction and benefit the network reconfiguration. Because of the nonconvexity of the three-phase balanced optimal power flow, a second-order cone program (SOCP) based approach is used to relax the optimal power flow problem. Then, the alternating direction method of multipliers (ADMM) is used to compute the optimal power flow in distributed manner. Considering the limited number of the switches and the increasing computation capability, the proposed network reconfiguration is solved in a parallel way. The numerical results demonstrate the feasible and effectiveness of the proposed approach.
\end{abstract}


\section*{Nomenclature}

\begin{notationlist}{MAXIMUMSPC}
\item[$\mathcal{G}$] The graph for PDS with a node set and a link set: $\mathcal{G} = [\mathcal{V}, \mathcal{E}]$.
	
\item[$\omega$] $\omega$ is a linear combination coefficient of SVR, $\frac{1}{2} \omega^T \omega$ indicates the flatness of the regression coefficients.

\item[$b$] $b$ is an offset coefficient of SVR.

\item[$C$] $C$ is a trade-off parameter of SVR. 

\item[$\gamma$] $\gamma$ is a parameter of (Gaussian) radial basis function.

\item[$\varepsilon$] $\varepsilon$ is an adjustable precision parameter, which indicates the training error threshold.

\item[$R_{risk}$] The objective function of designed for the SVR.
	
\item[$\varphi_1, \varphi_2$] Two positive acceleration coefficients for velocity updates of each particle in the PSO.

\item[$\theta_1$, $\theta_2$] Two independently random variables with uniformly distributed range $(0,1)$ for velocity updates of each particle in the PSO. 

\item[$\eta^{\Omega}_g$] The best position vector among all particles in PSO. 

\item[$\boldsymbol{\alpha}^{\Omega}_{i_3}$] The vector of position of the $i_3$th particle in PSO. 

\item[$\boldsymbol{\nu}^{\Omega}_{i_3}$] The vector of velocity of the $i_3$th particle in PSO.

\item[$\boldsymbol{\eta}^{\Omega}_{i_3}$] The vector of the best historical position of the $i_3$th particle in PSO.

\item[$V_i$] The complex voltage on bus $i$.

\item[$I_i$] The complex current to bus $i$.

\item[$z_{ij}$]	The complex impedance between bus $i$ and bus $j$, $z_{ij} = r_{ij} + \textbf{i} x_{ij}$.

\item[$S_{ij}$] The branch flow $S_{ij} = P_{ij} + \textbf{i} Q_{ij}$.

\item[$P_{ij}$] The real power $P_{ij}$ = $|I_{ij}|^2 r_{ij}$.

\item[$s_i$] Power injection on bus $i$.
\end{notationlist}

\section{Introduction}
\hspace*{3mm}

The network reconfiguration is a convenient way to reduce the system loss, maintain the voltage profiles, and balance the system load by opening and closing the limited number of switches~\cite{ding201645automatic,luo2016rea23l,bai2016in32terval,fang2016coup12on}. The developing load forecasting technology can provide the load deviations in the periods between two scheduled time points, which is usually ignored by traditional network reconfiguration approaches. Considering the stochastic deviation of load profiles in distribution system, it is imperative to build a dynamic and efficient network reconfiguration approach~\cite{zhang2017consumpti123on,c89ui2015wind, yigu2014statistical1, cui2017da123ta,zhu2016probabifewlistic,chen2016no123}. In this paper, a support vector regression (SVR) based short-term load forecasting approach is employed to cooperate with the network reconfiguration to minimize the system loss. 

Although most parameters in the SVR can be solved in a convex manner, several parameters (can be defined as hyper-parameters) cannot be determined by a similar way.  The optimization of these hyper-parameters is an indispensable issue in SVR, which dramatically impacts the efficiency and performance of the forecaster~\cite{burges1998tutorial, jiang2016shorfweft, chen2014rea123l,jian32g2017day}. Specifically, a two-step parameters optimization approach is proposed with grid traverse algorithm (GTA) and particle swarm optimization (PSO). Considering the distributed computation frameworks in~\cite{zaharia2010spa90rk,gu2016kno12wledge,yandistri2017buted,jiang2014faultbigdata}, the hyper-parameter optimization of SVR is designed based on "Mapreduce" idea to optimize the parameters in parallel and reduce the computation time in this paper.

In this paper, we are focusing on the three-phase balanced distribution system, and many heuristic methods are used to solve this problem. However, they are easily trapped into local minimums and heavily impacted by the initial points~\cite{tang2010globallypso,yigu2014statistical1,he201wef5joint}. To overcome this shortage, a distributed network reconfiguration approach is proposed for the distribution systems with convex relaxation. Specifically, a second-order cone program (SOCP) based approach is used to relax the three-phase balanced power flow problem~\cite{pe78ng2014distributed,jiang2014synchrophasor}. Then, the alternating direction methods of multipliers (ADMM) provides a distributed method to solving the relaxed optimal power flow problem~\cite{low2014co78nvex}. With the decomposition and dual ascent alternating, the ADMM can solve the relaxed network reconfiguration problem with a quickly convergence speed. Considering the increasing computation capability, a parallel searching manner is designed to traverse all the permutations of the switches, and determine the best configuration to minimize the system loss.

The paper is organized as follows. In Section~\ref{sec:formulation}, the flowchart of the proposed approach is introduced. In Section~\ref{Sec:SVR}, the SVR based approach is used to short-term forecast the distribution loads. In Section~\ref{sec:PDS}, based on the branch flow model, the OPF problem is relaxed with SOCP and solved with ADMM.  In Section~\ref{sec:results}, the numerical results are presented to validate the proposed approach. The conclusion is presented in Section~\ref{Sec:Concl}.

\section{The Architecture of the Proposed Approach}~\label{sec:formulation}  
\begin{figure*}[t]
	\begin{center}
		\includegraphics [width=1.155\columnwidth, angle=90]{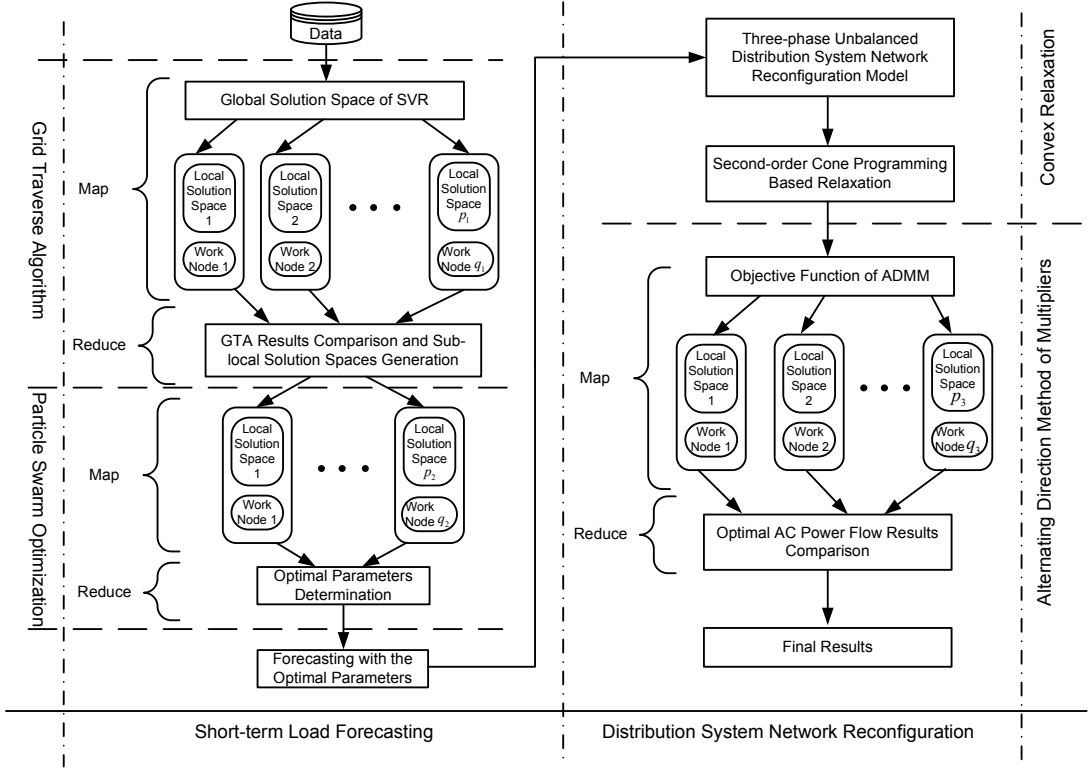}
		\caption{The flowchart of the proposed approach.}
		\label{fig:flowchart}
	\end{center}
\end{figure*}

As shown in Fig.~\ref{fig:flowchart}, the proposed approach consists of two parts, the short-term load forecasting and distribution system network reconfiguration. Both of them are designed in distributed manner to effectively organize computation resources and reduce the time consumption. 

As shown in the left part of Fig.~\ref{fig:flowchart}, in the part of short-term load forecasting, the optimization of the hyper-parameters contains GTA and PSO to avoid be trapped into local minimum. In the first step, the global solution space is spitted by the GTA into the local solution spaces. In the Map phase, because of the independency among the local solution spaces, they are traversed by GTA in parallel. In the Reduce phase, one or several local solution spaces are selected with the minimum training errors. In the second step, the selected local solution spaces are optimized by the PSO in the similar manner. Then, the optimal parameters can be generated after comparison in the Reduce phase. Finally, the short-term load can be forecasted with the optimized hyper-parameters. 

As shown in the right part of Fig.~\ref{fig:flowchart}, in the part of distribution system network reconfiguration, a three-phase balanced distribution system model is built with the forecasted load profiles. The SOCP is used to relax the three-phase balanced optimal power flow problem into a convex problem. In the Map phase, the ADMM is used to compute the three-phase balanced optimal power flow in parallel. Considering the very high speed of ADMM, in the Reduce phase, all the statuses of the switches are traversed and compared to get the global optimization results. Considering the limited number of the switches in the distribution system, all the permutations can be traversed in a parallel manner to determine the optimal configuration. Finally, the system loss can be reduced with the optimized distribution reconfiguration results.

\section{SVR-Based Short-term Load Forecaster}~\label{Sec:SVR}
\subsection{SVR Formulation}
In this part, the SVR-based short-term load forecaster is trained by the collected historical data to get the optimal hyper-parameters. The objective function of the Kernel based SVR can be built to minimize the forecast error with the soft margin as follows:
\begin{equation}
R_{risk} = \min\limits_{\varepsilon, \omega, \xi_{i_1}, \xi^*_{i_1}, C, b,\gamma} \{\frac{1}{2} \omega^T \omega + C\sum\limits_{{i_1} =1}^{n}(\xi_{i_1} + \xi^*_{i_1})\} 
\label{SVR_basic_risk2}
\end{equation} 
Subject to
\begin{equation}
\left\{
\begin{aligned}
&L'_{i_1} - f(x_{i_1}) \leqslant \varepsilon+\xi_{i_1}, \\
&-L'_{i_1} + f(x_{i_1}) \leqslant \varepsilon+\xi^*_{i_1},  \\
&\xi_{i_1}, \xi^*_{i_1} \geqslant 0.
\end{aligned}
\right.
\label{SVR_constrain}
\end{equation}
where in~(\ref{SVR_basic_risk2}),$f$ is a Kernel based regression function, $i_1$ is a time index, $\frac{1}{2} \omega^T \omega$ indicates the flatness of the regression coefficients, the second item is the tube violation, $C$ is a trade-off coefficient between the first two items, $\xi_{i_1}$ and $\xi^*_{i_1}$ indicates the two training errors. In~\cite{burges1998tutorial}, the risk function~(\ref{SVR_basic_risk2}) with the constraint~(\ref{SVR_constrain}) can be derived to a dual problem with Karush-Kuhn-Tucker (KKT) condition. However, the parameters $\gamma$, $C$, and $\varepsilon$ are still need to be determined, which are the critical factors to the performance of the forecaster~\cite{chang2011libsvm, burges1998tutorial}. The detail derivative of the SVM or SVR and its dual forms can be found in~\cite{chang2011libsvm, burges1998tutorial}. Then, as shown in Fig.~\ref{fig:flowchart}, a two-step based parameter optimization approach is designed to compute the optimal parameters. 
\subsection{Two-step parameter optimization}
Because the parameters $\gamma$, $C$, and $\varepsilon$ cannot be solved with the convex optimization, they are be defined as hyper-parameters in~\cite{bergstra2012ra23ndom}. There are several approaches are proposed for the hyper-parameters such as random search, sequential search, and Gaussian process~\cite{bergstra2012ra23ndom, bergstra2011algor23ithms}. Considering the complexity and feasibility, a grid traverse search based two-step hyper-parameter optimization is proposed for the SVR based short-term load forecasting~\cite{jiang2016shorfweft}.

\subsubsection{First Step: the GTA Procedure}\label{APP:4}
As shown in Fig.~\ref{fig:flowchart}, the GTA procedure is the first step for the hyper-parameters optimization, which aims to traverse the global solution space into one or several local solution spaces. In the second step, the local solution spaces can be searched with the PSO based approach. The proposed approach is based on the increasing computation capability and new computer cluster cooperation soughs, for example Mapreduce. In the first step, the three hyper-parameters are initialized with their upper bounds, lower bounds, and grid searching steps. Then, a traversing vector $\mathbf{H}$ can be generated as a finite multi-Cartesian product, which is critical for the Mapreduce process. $H_{j_2}$ is an element in $\mathbf{H}$. For each $H_{j_2}$, the loss function of SVR $R_{risk}$ can be computed independently. As shown in Fig.~\ref{fig:flowchart}, they can be computed in parallel to reduce the computation time. In the last step, the minimum $R_{risk}$ is selected. In addition, if several $H_{j_2}$ are selected, all of them are transmitted to the second step for PSO optimization. In real-applications, if there is high requirement for the time consumption, the GTA can provide good enough parameter optimization. 
\begin{algorithm}
	\caption{GTA for Hyper-parameter Optimization}
	\label{alg1}
	\begin{algorithmic}
		\STATE $\mathbf{Objective}$: Shrink the global solution spaces into one or several local solution spaces.
		\STATE
		\STATE $\mathbf{Initialization}$: Hyper-parameters initialization and multi-Cartesian product generation for the GTA.
		\STATE 
		\STATE $\mathbf{Grid\ Traverse\ Searching}$: For each core or each process, assign the $H_{j_2}$ to compute the $R_{risk}$, which can be computed in parallel with the Mapreduce model.
		\STATE
		\STATE $\mathbf{Determine\ Local\ Solution\ Space}$: Collected all the results, and select the local solution spaces with minimum $R_{risk}$.
	\end{algorithmic}
\end{algorithm}

\subsubsection{Second Step: the PSO Procedure}\label{APP:5}
In this paper, the PSO procedure is designed as a ``fine" optimization for the hyper-parameters, which can be implemented as the scenarios with less time consumption requirements.

After initialization of the particles, for particle $i_3$, its velocity and position can be computed as following:
\begin{subequations}\label{equ:constraint_PDS_BFM}
	\begin{align} 
	\boldsymbol{\nu}^{\Omega}_{i_3} (t) =& \boldsymbol{\nu}^{\Omega}_{i_3}(t-1)+\varphi_1\theta_1(\boldsymbol{\eta}^{\Omega}_{i_3} - \boldsymbol{\alpha}^{\Omega}_{i_3}(t-1)) \\ \nonumber
	&+ \varphi_2\theta_2(\boldsymbol{\eta}^{\Omega}_g - \boldsymbol{\alpha}^{\Omega}_{i_3}(t-1)), \\
	\boldsymbol{\alpha}^{\Omega}_{i_3} (t) =& \boldsymbol{\alpha}^{\Omega}_{i_3}(t-1)+\boldsymbol{\nu}^{\Omega}_{i_3}(t),
	\end{align}
\end{subequations} 
where the acceleration coefficients are defined as $\varphi_1$ and $\varphi_2$, $\theta_1$ and $\theta_2$  can be seemed as two independently weightiness coefficients, the best historical position and the best position are defined as $\boldsymbol{\eta}^{\Omega}_{i_3}$ and $\boldsymbol{\eta}^{\Omega}_g$, respectively, and hyper-parameter $\Omega$ = $[\gamma\  C\  \varepsilon]$. $\boldsymbol{\alpha}^{\Omega}_{i_3} (t)$ and $\boldsymbol{\nu}^{\Omega}_{i_3}(t)$ are the position and velocity vectors, respectively.

\section{Distribution System Network Reconfiguration}~\label{sec:PDS}
\begin{figure*}[!t]
	\begin{center}
		\subfigure[]{ \label{fig:IEEE 123}
			\resizebox{3.4in}{!}{\includegraphics{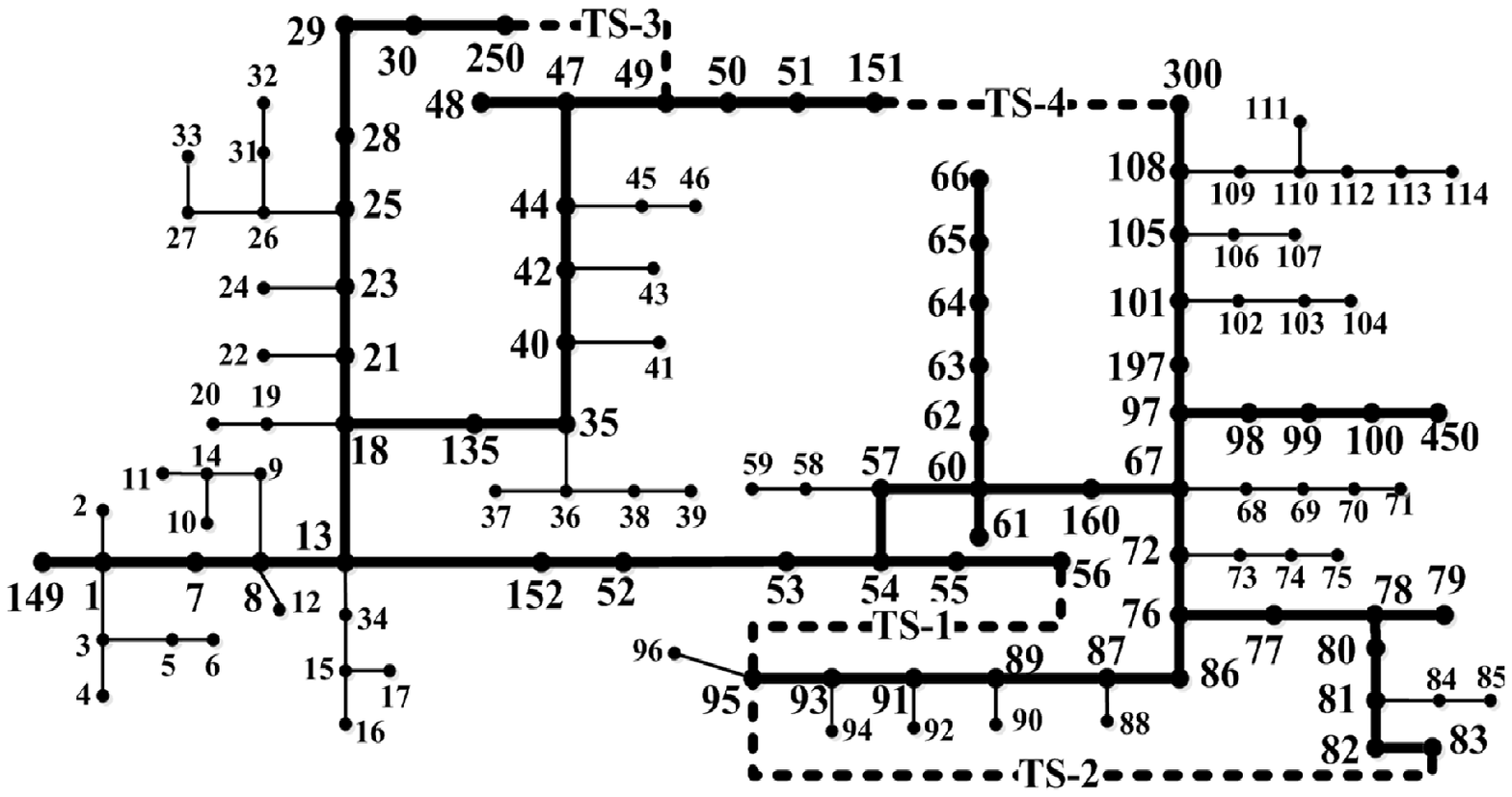}}}
		\subfigure[]{ \label{fig:Error Distri}
			\resizebox{3.4in}{!}{\includegraphics{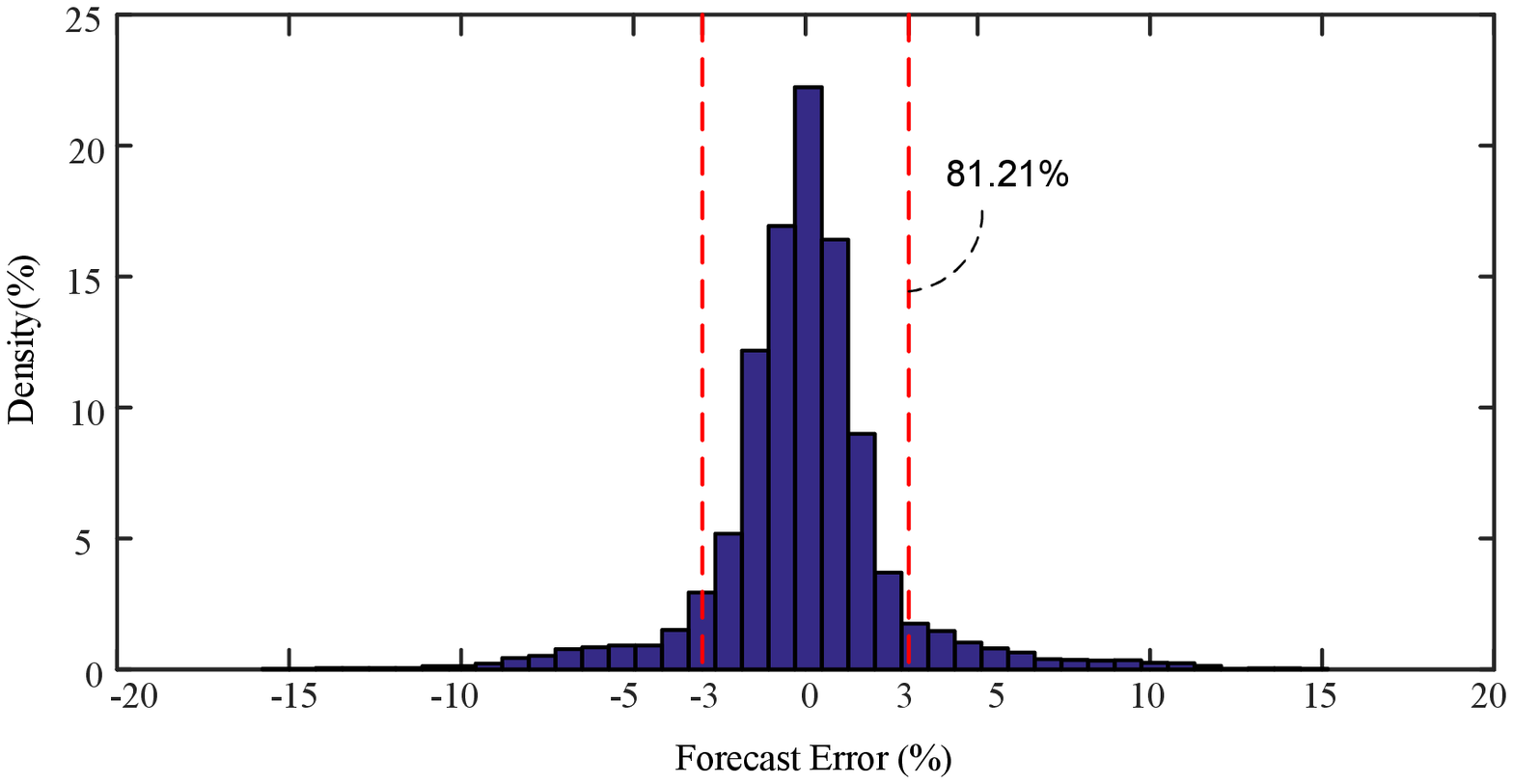}}}
		\caption{(a) The IEEE 123-bus based distribution system, (b) the error distribution of short-term load forecasting.}\label{fig:signal-analysis}
	\end{center}
\end{figure*}
\begin{center}
	\begin{table*}
		\caption{Results of Network Reconfiguration}\label{Table: basic}\vspace{-0.3cm}
		\begin{center}
			\begin{tabular}{ |c| c | c | l | c | c | c |}
				
				\hline
				No.&Scenario & Bus No. & Opened Switches & Original $P_{Loss}$  & New $P_{Loss}$ & \textbf{Loss Reduction}\\ \hline
				1.& Load Increasing & 83 &  TS-3, TS-4 & 54.2 Kw & 36.7 Kw & 32.28 \%\\ \hline
				2.& Load Increasing & 300 &  TS-2, TS-3 & 42.4 Kw & 31.5 Kw & 25.71 \%\\ \hline
				3.& Load Increasing & 95 &  TS-3, TS-4 & 78.5 Kw & 67.0 Kw & 14.65 \%\\ \hline
				4.& Load Increasing & 49 &  TS-1, TS-4 & 17.6 Kw & 13.4 Kw & 23.86 \%\\ \hline
				5.& Load Decreasing & 47 & TS-2, TS-4 & 39.4 Kw  & 22.3 Kw & 43.40 \% \\ \hline
				6.& Load Decreasing & 108 & TS-1, TS-4 & 24.1 Kw  & 20.1 Kw & 16.59 \% \\ \hline
				7.& Load Decreasing & 250 & TS-3, TS-4 & 29.5 Kw  & 26.8 Kw & 9.15 \% \\ \hline
				8.& Load Decreasing & 56 & TS-1, TS-2 & 35.2 Kw  & 33.7 Kw & 4.27 \% \\ \hline
			\end{tabular}
		\end{center}
	\end{table*}
\end{center}
The topology of a distribution system can be represented in a graph with buses and branches: $\mathcal{G} = [\mathcal{V}, \mathcal{E}]$. Then the branch flow model can be built as follows~\cite{pe78ng2014distributed,low2014co78nvex}:
\begin{subequations}
	\begin{align} 
	s_i &=  \sum_{j}S_{ij}-\sum_{k}(S_{ki}-l_{ki}z_{ki}),  \label{equ:eps1} \\
	v_j &= v_i-2(r_{ij}P_{ij} + x_{ij}Q_{ij}) + (r_{ij}^2 + x_{ij}^2)l_{ij}, \label{equ:eps2} \\
	l_{ij} &= (P^2_{ij} + Q^2_{ij})/v_j, \label{equ:eps3}
	\end{align}
\end{subequations}
where $l_{ij} := |I_{ij}|^2 $, $v_i:=|V_i|^2$, $S_{ij}$, $P_{ij}$, $Q_{ij}$ and $z_{ij}$ indicate the complex power flow, active power, reactive power, and impedance on branch $ij \in \mathcal{E}$, $S_{ij} = P_{ij} + \textbf{i} Q_{ij}$, $P_{ij}$ = $|I_{ij}|^2 r_{ij}$ and $z_{ij} = r_{ij} + \textbf{i} x_{ij}$.

During the operation of network reconfigurations, the topology of the distribution system is keeping radial and avoid any loops. Considering the characteristics of the three-phase balanced distribution system, the SOCP relaxation inequalities can be represented as follows~\cite{pe78ng2014distributed,low2014co78nvex}:
\begin{equation}
\frac{|S_{ij}|^2}{v_i} \leq l_{ij}, \label{equ:eps4}
\end{equation}
where (\ref{equ:eps4}) can be used to instead of (\ref{equ:eps3}) as the inequalities constraints. In this paper, the objective function is defined as total line loss as follows:
\begin{equation}
F = \sum_{\mathcal{E}}P_{ij}, \label{equ:eps5}
\end{equation} 
where the constraints contain (\ref{equ:eps1}), (\ref{equ:eps2}), (\ref{equ:eps4}), and the basic physical constraints:
\begin{subequations}
	\begin{align} 
	V_{i,min} \leq V_i \leq V_{i,max}, \label{equ:eps6} \\
	I_{ij} \leq I_{ij,max}. \label{equ:eps7} 
	\end{align}
\end{subequations}

Considering the ADMM, the objective function (\ref{equ:eps5}) with the constraints (\ref{equ:eps1}), (\ref{equ:eps2}), (\ref{equ:eps4}), (\ref{equ:eps6}), and (\ref{equ:eps7}) can be decomposed into a dual problem. The detail derivatives of the ADMM can be found in~\cite{pe78ng2014distributed,low2014co78nvex}.

During the parallel traverse of all statuses of the switches, the topology of the distribution system is keeping as a radial network without any loops, which can be formulated as~\cite{ding201645automatic}:
\begin{subequations}
	\begin{align} 
	\mathbf{rank}(A) = N-d, \label{equ:eps8} \\
	\sum_{\mathcal{E}} a_{ij} = N-d, \label{equ:eps9} \\
	\sum_{\mathcal{E}_k} a_{ij} = M_k-1. \label{equ:eps10}
	\end{align}
\end{subequations}
where $A$ is the adjacency matrix of the graph $\mathcal{G}$, $d$ is the number of slack bus, $N$ is the number of buses $\mathcal{V}$, $M_k$ is the number of branches in path $\mathcal{E}_k$, and $a_{ij}$ is an element of $A$:
\begin{equation}
a_{ij}=\left\{
\begin{aligned}
&1, \ if\  bus \ i \ and\  bus\  j\  are\  connected,\\
&0, \ else.
\end{aligned}
\right.
\label{equ:eps11}
\end{equation}
Considering the limited number of switches, the proposed approach is designed to traverse all the permutations and determine the optimal configuration of the distribution system. For example, the modified IEEE 123-bus system with 4 switches indicates 16 scenarios with all the permutations of the switches~\cite{ding201645automatic}. With the topology constraints discussed above, the number of scenarios can be reduced in different scenarios. Then, considering the independency of each configuration (permutation), all the permutation can be implemented into different cores or processes and computed independently, which dramatically reduces the computation time and keeps the convexity to get the guaranteed optimization results.

\section{Numerical Results}~\label{sec:results}
 As shown in Fig.~\ref{fig:IEEE 123}, the test bench is based on the IEEE 123-bus distribution system,  and four initially opened tie switches TS-1, TS-2, TS-3 and TS-4 are added to make the system topology changeable, and the detail information can be found in~\cite{ding201645automatic,ding2015hier78archical}. A test load data set is from a partner utility's distribution feeder.
\subsection{Short-term Load Forecasting}
The test load data contains four seasons of one year, and 30 days are selected for each season. The proposed SVR based load forecasting approach is used for 1-hour-ahead sliding window forecasting with 1 second resolution. The training data is 5 times as the test data. The distribution of the forecaster errors is shown in Fig.~\ref{fig:Error Distri}. The mean absolute precentage error (MAPE) is 2.23\%, normalized root-mean-square error (NRMSE) is 4.03\%, and more than 80\% of the errors are accumulated between (-3.1\%, 3.1\%). 
\subsection{Network Reconfiguration}
As shown in Table~\ref{Table: basic}, considering the load increasing, for example, in scenario 1, with the forecasting results, there is a load increasing 20.31\% in bus 83. The system loss reduces 32.28\% with the proposed approach. For the load decreasing, for example, in scenario 5, with the forecasting results, there is a load decreasing 55.4\% in bus 47. The system loss reduces 43.40\% with the proposed approach. The average loss reduction for the load increasing scenarios is 24.13\%, the average loss reduction for the load decreasing scenarios is 18.35\%, and the total average for all scenarios is 21.24\%.
\subsection{Comparison}
As shown in Table~\ref{Table: compare}, compared with the traditional network reconfiguration approach with the genetic algorithm (GA), the proposed approach has less computation time and more loss reduction. Furthermore, the proposed approach is more intuitive, and convenient for implementation in different programming language such as python and Matlab.  
\begin{center}
	\begin{table}
		\caption{Performance comparison}\label{Table: compare}\vspace{-0.3cm}
		\begin{center}
			\begin{tabular}{ |c| c | c | }
				
				\hline
				Methods& Loss Reduction & Computation Time (s) \\ \hline
				GA based traditional & 17.87\% & 107  \\ \hline
				Proposed Approach& 21.24\% & 30 \\ \hline
			\end{tabular}
		\end{center}
	\end{table}
\end{center}

\section{Conclusion}~\label{Sec:Concl}
In this paper, a short-term load forecasting based network reconfiguration is proposed to reduce the distribution system loss dynamically. Instead of the static load measurements at the scheduled time spots, the short-term load forecasting approach can provide the accurate future load profiles, which contains more information for the network reconfiguration. With the SOCP, the OPF of the three-phase balanced distribution system can be relaxed into a convex problem and solved with ADMM. The optimal reconfiguration is generated by the parallel computation with traversing all the permutations of the switche status. The whole proposed approach is designed as a distributed computation approach.

In the next step, with the increasing penetration of renewable energies, the wind and solar power will be considered, and other system such as traffic system with electrical vehicles will be also considered in the future research.


\bibliographystyle{IEEEtran}
\bibliography{IEEEfull,refs_liu_2017_asilomar}
\end{document}